# BESS Siting and Sizing to Mitigate EV Charging-Induced Voltage Violations: A Planning Space Reduction Approach

Linhan Fang, *Student Member, IEEE* and Xingpeng Li, *Senior Member, IEEE*

***Abstract*—The rapid growth of electric vehicle (EV) charging can cause persistent voltage violations in distribution networks and increase the need for grid-side voltage support. This paper develops a proactive voltage management framework for planning battery energy storage systems (BESS) to mitigate EV charging-induced voltage violations. Annual EV charging profiles are generated through Monte Carlo simulation using probability distributions derived from real charging events and are incorporated into an annual voltage assessment to identify stressed operating conditions. An optimal expansion planning model then determines installation locations and capacities at minimum investment cost. To improve computational tractability, a planning space reduction (PSR) strategy combines critical period selection to reduce the chronological horizon with candidate bus screening to reduce the BESS siting space. The resulting investment decisions are evaluated through full-year operational validation on the IEEE 33-bus, IEEE 69-bus, and a 240-bus distribution system. Results show that PSR reduces solution time by up to 99.2% while matching the exhaustive benchmark in the 33-bus and 69-bus systems and limiting the objective deviation to 0.55% in the 240-bus system. The optimized BESS deployments eliminate the identified annual voltage violations and further reduce electricity purchase costs and network energy losses under increasing EV penetration.**

***Index Terms*—Battery energy storage system, BESS siting and sizing, Distribution system planning, Electric vehicle charging, Planning space reduction, Voltage violation mitigation.**

## NOMENCLATURE

*A. Indices and Sets*

| | |
|---|---|
| $s \in \mathcal{G}$ | Index and set of slack buses |
| $i \in \mathcal{N}$ | Index and set of non-slack buses |
| $(i,j) \in \mathcal{L}$ | Index and set of branch information |
| $t \in \mathcal{T}$ | Index and set of time intervals |
| $j \in \mathcal{D}(i)$ | Index and set of downstream nodes of node $i$ |
| $k \in \mathcal{U}(i)$ | Index and set of upstream nodes of node $i$ |
| $b \in \mathcal{B}$ | Index and set of candidate BESS installation buses |
| $r \in D_d$ | Index and set of violation timestamps on date $d$ |
| $b \in c$ | Index and set of candidate buses in each cluster |

*B. Parameters*

| | |
|---|---|
| $r_{ki}$ | Resistance of the line connecting node $k$ and node $i$ |
| $x_{ki}$ | Reactance of the line connecting node $k$ and node $i$ |
| $P_{load,i}^{t}$ | Active power demand of node $i$ at time $t$ |
| $Q_{load,i}^{t}$ | Reactive power demand of node $i$ at time $t$ |
| $E_{min}^{cap}, E_{max}^{cap}$ | Minimum / maximum installable capacity of BESS |
| $SoC^{min}$ | Minimum limit of state of charge |
| $SoC^{max}$ | Maximum limit of state of charge |
| $\eta_{charge}$ | Charging power efficiency of BESS |
| $\eta_{discharge}$ | Discharging power efficiency of BESS |
| $c^{cap}$ | Installation capacity cost of BESS |
| $c_t^{operation}$ | Operational cost at time $t$ |
| $V_L, V_U$ | Voltage lower limit / upper limit |
| $C_{rate}^{ch}, C_{rate}^{dis}$ | Charge / discharge C-rate of BESS |
| $K_Q^{inj}, K_Q^{abs}$ | Injected / absorbed reactive power-to-energy capacity ratio of BESS |

*C. Variables*

| | |
|---|---|
| $P_{ij}^t, Q_{ij}^t$ | Power flow from node $i$ to node $j$ at time $t$ |
| $P_s^t, Q_s^t$ | Total power consumption of the system at time $t$ |
| $v_i^t$ | The squared of voltage magnitude of node $i$ at $t$ |
| $V_s^t$ | Time-varying average voltage at slack bus at time $t$ |
| $l_{ij}^t$ | The square of the current in the line at time $t$ |
| $P_{b,t}^{BESS_ch}$ | Active charging power of BESS at bus $b$ at time $t$ |
| $P_{b,t}^{BESS_dis}$ | Active discharging power of BESS at bus $b$ at time $t$ |
| $Q_{b,t}^{BESS_inj}$ | Reactive power flowing from BESS into grid at bus $b$ at time $t$ |
| $Q_{b,t}^{BESS_abs}$ | Reactive power flowing from grid into BESS at bus $b$ at time $t$ |
| $SoC_{b,t}$ | The state of charge of BESS at bus $b$ at time $t$ |
| $E_b^{cap}$ | The installed capacity of BESS at bus $b$ |
| $E_{b,t}^{ess}$ | The amount of the stored energy of BESS at bus $b$ at time $t$ |
| $S_r$ | Voltage severity for the violation at timestamp $r$ |
| $V_r$ | Actual voltage value associated with the violation record at timestamp $r$ |
| $M_{count,d}$ | Number of violations on date $d$ |
| $M_{total_sev,d}$ | Total violations severity on date $d$ |
| $M_{max_sev,d}$ | Maximum violations severity on date $d$ |
| $M_{duration,d}$ | Violations duration proxy on date $d$ |
| $S_{mean_abs,b}$ | Mean absolute sensitivity of the bus $b$ |
| $S_{k,b}$ | Voltage sensitivity of bus $b$ to the $k$th disturbance |
| $F_{viol,b}$ | Total violation frequency of bus $b$ |
| $P_{UV,b}, P_{OV,b}$ | Undervoltage / overvoltage time step percentages of bus $b$ |
| $S_{eol,b}$ | End of line score of bus $b$ |

## I. INTRODUCTION

The electrification of transportation is introducing a rapidly growing source of demand to distribution networks, with the resulting grid impacts depending strongly on charging power, location, and temporal coincidence with existing loads. Residential Level-2 charging is particularly relevant because it provides substantially higher charging capability than Level-1 charging while avoiding the infrastructure requirements of DC fast charging [1]-[2]. As adoption increases, the aggregation of

Linhan Fang and Xingpeng Li are with the Department of Electrical and Computer Engineering, University of Houston, Houston, TX, 77204, USA. (e-mail: lfang7@uh.edu; xli83@central.uh.edu).

Acknowledgment: We thank Center Point Energy for providing real smart meter data and for their feedback on method for extracting EV charging loads.

these charging loads can intensify feeder loading, deteriorate voltage profiles, and accelerate the utilization and aging of distribution assets, creating additional reliability and planning concerns [3]-[4]. At the same time, EVs are becoming an increasingly important component of the broader energy system, with the potential to support grid operation and facilitate renewable energy integration when their flexibility is appropriately coordinated [5]-[7]. These characteristics create a need for distribution planning approaches that can accommodate continued EV growth while maintaining acceptable network operating conditions and limiting unnecessary infrastructure investment.

In literature, data-driven approaches have gained traction due to their ability to operate without precise system models [8]-[10], [15]. For instance, [9] proposes a model predictive control-based method using piecewise linear regression to regulate voltage and power in active distribution networks. Also, the work in [8] introduces a distributed data-driven optimization framework combining recursive kernel regression and alternating direction method of multipliers (ADMM), enabling rapid response to system changes. However, the reliability of data-driven approaches is fundamentally compromised by the volatile nature of distributed energy resources (DERs) and loads. This issue is further compounded by the practical difficulty of acquiring sufficient, high-fidelity data, particularly for complex, large-scale power system testbeds.

Deep learning and reinforcement learning techniques also have emerged as powerful tools for voltage regulation [11]-[14]. The integration of machine learning with traditional optimization models presents a powerful paradigm for solving complex problems, offering significant improvements in both computational efficiency and solution quality. [11] proposes a convolutional neural network-based stochastic distribution network reconfiguration method, optimizing topology to reduce power losses and enhance voltage stability. [12] develops a day-ahead multi-agent deep reinforcement learning framework for dynamic voltage regulation, leveraging smart inverters to minimize voltage deviations. [13] further advances this field with a graph-based multi-agent reinforcement learning method, enabling decentralized voltage regulation in multi-microgrid networks. [14] presents a two-timescale coordinated voltage regulation method using hierarchical multi-agent reinforcement learning.

The coordination of multiple hybrid energy resources, such as battery energy storage systems (BESS), capacitors, and photovoltaic (PV) systems, is critical for voltage regulation [15]-[18]. Reference [15] proposes a data-driven Volt-VAR scheduling strategy that leverages a mobile energy storage system for day-ahead voltage regulation in distribution networks with high penetrations of PV and wind power. The work in [16] puts forward a budget-constrained model that co-optimizes the siting, sizing, and operation of distributed BESS. The objective is to concurrently maximize revenue from ancillary services while enhancing the operational performance and reliability of unbalanced distribution networks. The work in [17] presents a realistic, linear model for BESS, highlighting the importance of capturing variable efficiency and nonlinear characteristics of BESS for accurate power system studies. [18] provides a comprehensive review of energy flexibility in modern power systems, emphasizing the critical role of coordinating flexible resources such as BESS to maintain grid stability and support the integration of renewable energy sources. These studies provide a robust theoretical foundation for multi-device coordination.

Determining the optimal location and capacity of BESS is an important research focus. [19] and [20] investigate the optimal placement and sizing of BESS, using heat map visualizations and optimization algorithms like particle swarm optimization to minimize power losses and voltage deviations. [21] proposes a multi-objective optimization model considering flexibility and economy, solved using the non-dominated sorting genetic algorithm-II. [22] presents a method for determining the minimum number and optimal locations of phasor measurement units (PMUs) to ensure complete system observability. When multiple optimal solutions are found, a robustness-based system observability redundancy index is utilized to rank these configurations. The proposed framework may also be applicable to the optimal BESS site selection problems.

Besides, the integration of PV and EV charging stations introduces new challenges. [23] proposes an optimal planning method incorporating PV-grid-EV transactions and a peer-to-peer market mechanism to enhance grid security. [2] explores the use of PV and BESS to mitigate electricity costs for fast EV charging, deploying the direct-current fast charging station in conjunction with PV panels and the BESS. Moreover, the practical acquisition and application of EV charging data are complicated by real-world constraints, including strict user privacy concerns and the inherent variability of charging power, which is highly dependent on the state of charge (SoC) [24]-[26]. Incorporating battery degradation presents a key challenge in optimization modeling [27]. On one hand, adding degradation-aware constraints is vital for ensuring economic evaluations are not overly optimistic. On the other hand, this process invariably leads to a substantial increase in model complexity and computational burden [28].

To summarize, there remain several challenges in solving the OEP model that aims to address the voltage violation issues by installing BESS. These challenges are as follows:

- The Long chronological horizons and large candidate sets can substantially increase the computational burden of BESS planning, leading to excessive computing time, or even rendering the OEP problem unsolvable.
- Although selecting all violation nodes as candidate locations for BESS installation makes OEP highly comprehensive, it can also cause a combinatorial explosion and exponentially increase the OEP model's computational complexity, which may lead to divergence issues.
- The manual selection of BESS installation candidate nodes and optimization model solution parameters introduces credibility issue, as these processes lack objectivity and convincing evidence.

This paper proposes a novel proactive voltage management (PVM) framework to address these challenges. It can efficiently solve the optimal BESS planning problem for mitigating voltage violations caused by realistic EV charging loads. The PVM framework begins by generating high-fidelity stochastic EV charging scenarios derived from real EV user data to ensure practical relevance. The proposed planning space reduction (PSR) strategy reduces the temporal planning horizon and BESS candidate space before optimization, thereby improving computational tractability. By identifying critical time periods for monitoring and candidate locations for BESS placement, the proposed PSR strategy makes the large-scale planning problem computationally tractable while retaining solution quality. The technical efficacy and economic benefits of the proposed OEP model in resolving voltage issues have been rigorously validated on multiple standard test systems, confirming both its rationality and feasibility. The contributions of this paper are presented as follows:

- A proactive voltage management (PVM) framework is developed to assess and mitigate EV charging-induced voltage violations in distribution networks. The framework integrates stochastic EV charging scenario generation, annual voltage violation assessment, BESS planning, and full year validation within a unified planning process.
- A VVA model is established to characterize annual voltage violations under increasing EV penetration. The planning study incorporates stochastic EV charging profiles generated from empirically derived charging-event distributions.
- A BESS planning model is formulated to jointly determine installation locations and capacities for mitigating EV induced voltage violations at minimum investment cost. The resulting BESS plans are further evaluated over the full annual horizon and assessed for their operational and economic benefits under different EV penetration levels.
- A planning space reduction (PSR) strategy is developed to improve the computational tractability of BESS planning. It combines critical period selection (CPS), which identifies high stress chronological periods from annual violation records, with candidate bus screening (CBS), which reduces the BESS siting space using voltage sensitivity, violation characteristics, clustering, and network information. Tests on the IEEE 33-bus, IEEE 69-bus, and 240-bus systems demonstrate substantial reductions in computational burden while maintaining comparable planning performance to the corresponding benchmark methods.

## II. The Proposed OEP Model for Optimal BESS Sizing and Placement

### *A. Voltage Violation Analysis (VVA) Model*

Annual voltage screening is performed with an hourly radial branch flow formulation [29]. Nodal voltages and branch flows are evaluated over the projected annual load trajectory, and intervals exceeding the prescribed voltage limits are recorded for subsequent BESS planning. The analysis is conducted at hourly resolution over the annual horizon in the VVA model.

The objective of the proposed VVA model is to minimize the total active power losses as obtained in (1).

$$objective = \min\left(\sum_{(i,j)\in\mathcal{L}}\sum_{t\in\mathcal{T}} r_{ij}\cdot l_{ij}^t\right) \tag{1}$$

where $\mathcal{L}$ represents the set of branches and $r_{ij}$ denotes the resistance of the branch connecting node $i$ and node $j$.

The nodal active and reactive power balance equations at non-slack and slack buses are expressed as (2)-(3) respectively. Slack buses represent the substation buses.

$$\begin{cases}\sum_{k\in\mathcal{U}(i)}(P_{ki}^t - r_{ki}\cdot l_{ki}^t) = \sum_{j\in\mathcal{D}(i)} P_{ij}^t + P_{load,i}^t \\ \sum_{k\in\mathcal{U}(i)}(Q_{ki}^t - x_{ki}\cdot l_{ki}^t) = \sum_{j\in\mathcal{D}(i)} Q_{ij}^t + Q_{load,i}^t \\ \forall i\in\mathcal{N}, \forall t\in\mathcal{T}\end{cases} \tag{2}$$

$$\begin{cases}\sum_{k\in\mathcal{U}(i)}(P_{ki}^t - r_{ki}\cdot l_{ki}^t) = -P_s^t + \sum_{j\in\mathcal{D}(i)} P_{ij}^t + P_{load,i}^t \\ \sum_{k\in\mathcal{U}(i)}(Q_{ki}^t - x_{ki}\cdot l_{ki}^t) = -Q_s^t + \sum_{j\in\mathcal{D}(i)} Q_{ij}^t + Q_{load,i}^t \\ \forall s\in\mathcal{G}, \forall t\in\mathcal{T}\end{cases} \tag{3}$$

where for each node $i\in\mathcal{N}$, $\mathcal{D}(i)$ and $\mathcal{U}(i)$ represent the sets of its downstream and upstream nodes, respectively.

The voltage drop equation can be obtained as (4) and the voltage constraint of the slack bus is obtained as (5).

$$v_i^t - v_j^t = 2(r_{ij}P_{ij}^t + x_{ij}Q_{ij}^t) - (r_{ij}^2 + x_{ij}^2)l_{ij}^t \quad ,\forall(i,j)\in\mathcal{L}, \forall t\in\mathcal{T} \tag{4}$$

$$v_s^t = (V_s^t)^2, \forall s\in\mathcal{G}, \forall t\in\mathcal{T} \tag{5}$$

The constraint between current and power (second-order cone constraint) is defined in (6).

$$v_i^t\cdot l_{ij}^t \geq (P_{ij}^t)^2 + (Q_{ij}^t)^2, \forall(i,j)\in\mathcal{L}, \forall t\in\mathcal{T} \tag{6}$$

The resulting annual violation records provide the temporal information for CPS and identify the buses considered in the subsequent CBS procedure, as detailed in Section III.

### *B. Optimal Expansion Planning (OEP) Model with BESS*

To mitigate the identified voltage violations, the OEP model is developed as a planning engine that co-optimizes BESS sizing and placement. The objective function of the OEP model, as defined below, is to minimize the BESS capital cost. While some fundamental constraints from VVA model are adopted with or without modifications, additional constraints related to BESS are included in the OEP model.

$$objective = \min\left(\sum_{b\in\mathcal{B}} c^{cap}\cdot E_b^{cap}\right) \tag{7}$$

where $c^{cap}$ represents the capacity cost of the BESS.

As enforced in (8)-(11), $z_b$ is a binary variable defining the BESS installation status at bus $b$. $u_{b,t}^{charge}$ and $u_{b,t}^{discharge}$ are binary variables indicating charging or discharging status respectively. $u_{b,t}^{inj}$ and $u_{b,t}^{abs}$ are binary variables indicating injection or absorption status respectively. Besides, $E_b^{cap}$ bounds the installed energy capacity. The proposed reduced OEP model monitors a subset of selected critical periods as the solution time $\mathcal{T}$.

$$z_b, u_{b,t}^{inj}, u_{b,t}^{abs}, u_{b,t}^{charge}, u_{b,t}^{discharge} \in \{0,1\} \tag{8}$$

$$z_b\cdot E_{min}^{cap} \leq E_b^{cap} \leq z_b\cdot E_{max}^{cap}, \forall b\in\mathcal{B} \tag{9}$$

$$u_{b,t}^{inj} + u_{b,t}^{abs} \leq 1, \forall b\in\mathcal{B}, \forall t\in\mathcal{T} \tag{10}$$

$$u_{b,t}^{charge} + u_{b,t}^{discharge} \leq 1, \forall b\in\mathcal{B}, \forall t\in\mathcal{T} \tag{11}$$

The mode binaries are linked to the active and reactive power variables through the Big-M constraints in (12)-(19), where $M$ is a large positive number.

$$0 \leq P_{b,t}^{BESS_ch} \leq C_{rate}^{ch}\cdot E_b^{cap}, \forall b\in\mathcal{B}, \forall t\in\mathcal{T} \tag{12}$$

$$0 \leq P_{b,t}^{BESS_ch} \leq M\cdot u_{b,t}^{charge}, \forall b\in\mathcal{B}, \forall t\in\mathcal{T} \tag{13}$$

$$0 \le P_{b,t}^{BESS_dis} \le C_{rate}^{dis} \cdot E_b^{cap}, \forall b \in \mathcal{B}, \forall t \in \mathcal{T} \quad (14)$$

$$0 \le P_{b,t}^{BESS_dis} \le M \cdot u_{b,t}^{discharge}, \forall b \in \mathcal{B}, \forall t \in \mathcal{T} \quad (15)$$

$$0 \le Q_{b,t}^{BESS_inj} \le K_Q^{inj} \cdot E_b^{cap}, \forall b \in \mathcal{B}, \forall t \in \mathcal{T} \quad (16)$$

$$0 \le Q_{b,t}^{BESS_inj} \le M \cdot u_{b,t}^{inj}, \forall b \in \mathcal{B}, \forall t \in \mathcal{T} \quad (17)$$

$$0 \le Q_{b,t}^{BESS_abs} \le K_Q^{abs} \cdot E_b^{cap}, \forall b \in \mathcal{B}, \forall t \in \mathcal{T} \quad (18)$$

$$0 \le Q_{b,t}^{BESS_abs} \le M \cdot u_{b,t}^{abs}, \forall b \in \mathcal{B}, \forall t \in \mathcal{T} \quad (19)$$

where the adopted $C_{rate}^{ch}$, $C_{rate}^{dis}$, $K_Q^{inj}$ and $K_Q^{abs}$ values are selected such that the combined active and reactive power limits remain within the apparent power rating for the inverter.

Equation (20) bounds the stored energy within the prescribed SoC range.

$$SoC^{min} \cdot E_b^{cap} \le E_{b,t}^{ess} \le SoC^{max} \cdot E_b^{cap}, \forall b \in \mathcal{B}, \forall t \in \mathcal{T} \quad (20)$$

Equation (21) updates the stored energy across consecutive intervals with charging and discharging efficiencies. Equation (22) imposes equal initial and terminal energy levels over the planning period at the beginning time $t_0$ and end time $t_{end}$.

$$E_{b,t}^{ess} = E_{b,t-1}^{ess} + \left( P_{b,t}^{BESS_ch} \eta_{charge} - \frac{P_{b,t}^{BESS_dis}}{\eta_{discharge}} \right) \Delta t, \forall b \in \mathcal{B} \quad (21)$$

$$E_{b,t_0}^{ess} = E_{b,t_{end}}^{ess}, \forall b \in \mathcal{B} \quad (22)$$

Nodal active and reactive power balance equations for non-slack nodes are defined in (23) and (24).

$$\sum_{k \in \mathcal{U}(i)} (P_{ki}^t - r_{ki} \cdot l_{ki}^t) = \sum_{j \in \mathcal{D}(i)} P_{ij}^t + P_{load,i}^t + \sum_{b \in \mathcal{B}_i} (P_{b,t}^{BESS_ch} - P_{b,t}^{BESS_dis}), \forall i \in \mathcal{N}, \forall t \in \mathcal{T} \quad (23)$$

$$\sum_{k \in \mathcal{U}(i)} (Q_{ki}^t - x_{ki} \cdot l_{ki}^t) = \sum_{j \in \mathcal{D}(i)} Q_{ij}^t + Q_{load,i}^t + \sum_{b \in \mathcal{B}_i} (Q_{b,t}^{BESS_abs} - Q_{b,t}^{BESS_inj}), \forall i \in \mathcal{N}, \forall t \in \mathcal{T} \quad (24)$$

The proposed OEP model determines the optimal BESS installation locations and the associated capacities to efficiently address the violation issues with a minimum cost. With the voltage violations reported from the VVA model, the proposed OEP model can operate on a reduced problem space of selected critical periods, rather than operating the OEP model throughout the entire year, where a large number of hourly periods have no impact on the optimal values of decision variables.

Table 1. Comparison of Key Settings between VVA model and OEP model.

| *Model* | *Objective* | *Constraints* | *Timeframe $\mathcal{T}$* |
|---|---|---|---|
| VVA model | (1) | (2)-(6) | Full year |
| Full OEP model | (7) | (3)-(6), (8)-(24) | Full year |
| Reduced OEP model | (7) | (3)-(6), (8)-(24) | Critical period |

The OEP model size reduction can be achieved with the proposed PSR strategy, including the CPS and CBS methods, which are presented in the next section. Table 1 summarizes the formulations for the proposed VVA and reduced OEP models.

## III. Critical Period and Candidate Bus Selection

The PVM framework is designed to manage large-scale optimization through the proposed PSR strategy, as shown in Fig. 1. It begins by using the VVA model as a preliminary screening tool. If VVA detects no voltage violations, the reduced OEP model for mitigation is deemed unnecessary. If violations are identified, PSR supplies the OEP model with reduced temporal and spatial input sets. The CPS method identifies a critical chronological period from annual voltage violation records, whereas the CBS method constructs a compact BESS candidate set from bus criticality and network information. The resulting planning solution is subsequently checked over the full annual horizon.

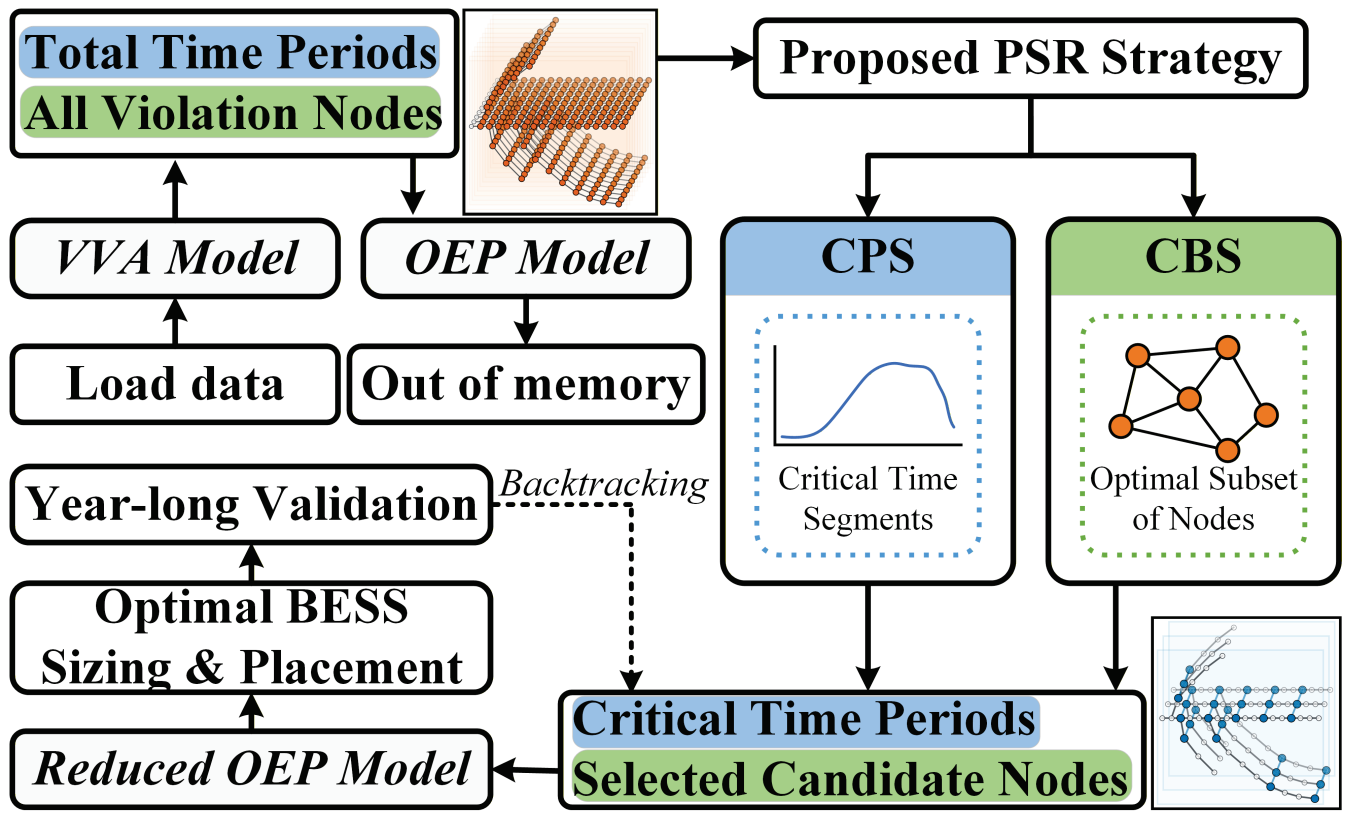

Fig. 1. Schematic diagram of the PVM framework using the PSR strategy.

If the optimal solution fails the year-long validation, the process will backtrack to the critical time period selection and then additional time periods will be adjusted to resolve the planning model. This proposed preprocessing strategy significantly enhances the computational efficiency and performance of the reduced OEP model.

### A. Critical Period Selection

The first stage of the PSR strategy is to assess the temporal criticality by identifying the most severe periods for optimization analysis as illustrated in Fig. 2. To evaluate and quantify voltage violations, we propose four metrics: total frequency, aggregate magnitude, peak magnitude, and cumulative duration. The representative critical time period for monitoring in the OEP model can then be determined using a weighted value of these metrics.

The voltage violation severity can be calculated as follows:

$$S_r = \begin{cases} V_L - V_r \ \ if \ V_r < V_L \\ V_r - V_U \ if \ V_r > V_U, r \in D_d \\ \quad 0 \quad \ others \end{cases} \quad (25)$$

where $S_r$ is the voltage severity at the violation timestamp $r$, and $D_d$ is the set of all recorded voltage violation event timestamps occurring on date $d$.

The number of violations, total violations severity, maximum violations severity, and violations duration proxy can be calculated using (26)-(29).

$$M_{count,d} = \textstyle\sum_{r \in D_d} 1 \quad (26)$$

$$M_{total_sev,d} = \textstyle\sum_{r \in D_d} S_r \quad (27)$$

$$M_{max_sev,d} = max_{r \in D_d} S_r \quad (28)$$

$$M_{duration,d} = \left| \textstyle\bigcup_{r \in D_d} H_r \right| \quad (29)$$

where $H_r$ is the non-repeated hour corresponding to the violation timestamp $r$.

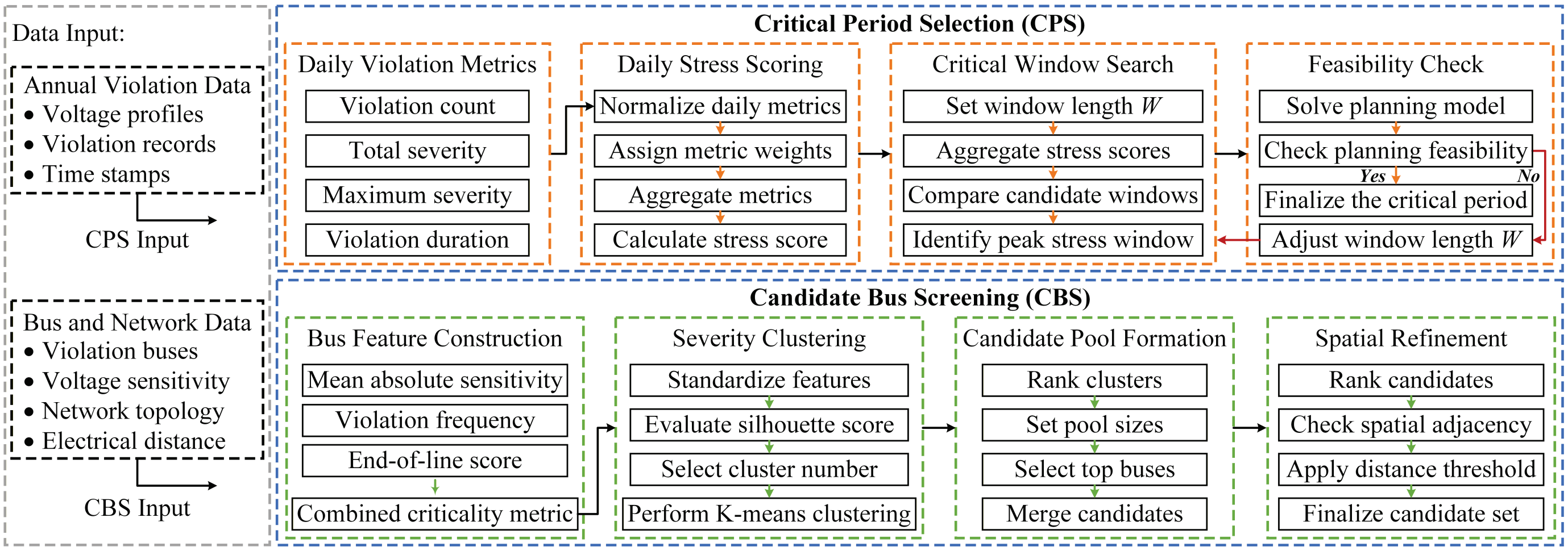

Fig. 2. Flowchart of the PSR strategy combining CPS and CBS methods.

To create a combined daily stress score, the severity indicators are integrated through a two-step process:

1) Metric normalization: each daily indicator $M_{k,d}$ consists of the above four indicators and timestamp. The normalization process of $M_{k,d}$ can be calculated with (30), where, $min_{d'}$ / $max_{d'}$ are the minimum and maximum value of the corresponding indicator for all dates:

$$M'_{k,d} = \frac{M_{k,d} - min_{d'}(M'_{k,d})}{max_{d'}(M'_{k,d}) - min_{d'}(M'_{k,d})} \tag{30}$$

2) Weighted combination: the weighted metric can be calculated using (31), where $\omega_k$ is the weight corresponding to indicator $k$:

$$S_{daily_stress,d} = \omega_{count} \cdot M'_{count,d} + \omega_{total_sev} \cdot M'_{total_sev,d} + \omega_{\max_sev} \cdot M'_{\max_sev,d} + \omega_{duration} \cdot M'_{duration,d} \tag{31}$$

Using a sliding window of $W$ days, the cumulative stress score for a period $p$ ending on date $d_{end}$ is calculated by (32) and the most severe time period is then identified as the maximum value obtained from (33).

$$R_p = \sum_{i=0}^{W-1} S_{daily_stress,(d_{end}-i)} \tag{32}$$

$$P^{worst} = arg\,max_p(R_p) \tag{33}$$

The start and end dates for the most severe period $P^{worst}$ are defined as ($d^*_{end} - W + 1$) and $d^*_{end}$, where $d^*_{end}$ corresponds to the date with the maximum cumulative stress score $R_{p^{worst}}$. This approach performs a planning analysis on the most challenging cumulative period under an energy-neutral constraint, thereby focusing the planning model on the most severe cumulative voltage-stress period.

## B. Candidate Bus Selection

The second stage of the strategy, the CBS method constructs a reduced set of BESS candidate buses from buses exhibiting voltage violations. Voltage sensitivity, violation frequency, clustering, and network-location information are used to form and refine the candidate set. It then applies a screening mechanism to adjust the node selection strategy based on the unique characteristics of the specific system. The whole process is illustrated in Fig. 2. The CBS method for candidate nodes selection is comprised of four phases:

Phase (i) - Feature Engineering for Node Criticality: voltage violation data, voltage sensitivity data, and electrical distance data are loaded, and the network topology graph $G = (\mathcal{N}, \mathcal{L})$ are constructed. To quantify the criticality of each node experiencing voltage violations, the average absolute sensitivity is determined. This is computed by simulating a 0.01 per unit active power injection at the specific node and then averaging the magnitudes of the resulting voltage sensitivities across the entire network. Then the mean absolute sensitivity is expressed as:

$$S_{mean_abs,b} = \frac{1}{N_k}\sum_k |S_{k,b}| \tag{34}$$

where $N_k$ is the number of sensitivity indicators and the violation frequency of each node is shown as:

$$F_{viol,b} = P_{UV,b} + P_{OV,b} \tag{35}$$

A topological end node is assigned a label value of 1 with $E_{topo,b} = 1$. This is used to calculate the end of line score $S_{eol,b} = \alpha_{eol} \cdot E_{topo,b}$. Subsequently, this score is integrated into the combined metrics defined in (36) to assess the potential criticality of a busbar for BESS installation. The same symbols are retained for the normalized values in (36).

$$M_{comb,b} = S_{maen_abs,b} + F_{viol,b} + S_{eol,b} \tag{36}$$

Phase (ii) - Severity Clustering: The characteristic data in obtained in the feature engineering phase are standardized here to obtain $X_{scaled}$ and then the Silhouette Score is used to evaluate the effect of different clustering numbers $K$ as:

$$K_{opt} = arg\,max_K\, Silhouette\,(X_{scaled}, K) \tag{37}$$

where $K_{opt}$ is the optimal number of clusters and the $X_{scaled}$ is subsequently processed using K-Means clustering with the $K_{opt}$. Through this procedure, each bus $b$ exhibiting voltage violations is assigned a distinct severity cluster label $C_b$.

Phase (iii) - Candidate Bus Selection: This phase will sort clusters in descending order according to the average comprehensive index of the busbars within each cluster $c$. The average comprehensive index can be obtained by:

$$\bar{M}_{comb,b} = \underset{b \in c}{\text{mean}}(M_{comb,b}) \tag{38}$$

For the $j$-th cluster $c_j$ after sorting, the number of nodes is:

$$N_{pool,c_j} = N_{max_top} - \left(\frac{N_{max_top} - N_{min_bottom}}{K_{opt} - 1}\right) \cdot j \tag{39}$$

Subsequently, the first $N_{pool,c_j}$ busbars are selected from each cluster $c_j$ based on the highest combined metric index $M_{comb,b}$. These selected busbars are then merged, and any duplicates are removed, to form the candidate pool $P_{start}$.

Phase (iv) - Spatial Diversity Adjustment: To refine the candidate nodes selection, the buses in $P_{start}$ are arranged in descending order according to the combined metric index $M_{comb,b}$. An iterative algorithm is then applied to this sorted list to construct the final, spatially diverse candidate set. The algorithm begins by selecting the top-ranked node and then processes the remaining nodes in the list. A subsequent node is added to the candidate set only if it satisfies a spatial separation criterion relative to all previously selected nodes. This criterion requires that a candidate node either: (i) is not topologically adjacent to any previously selected node, or (ii) if adjacent, its electrical distance from all previously selected nodes exceeds a predefined threshold. This iterative process continues until the target number of candidates is reached.

### *C. Full Year Validation*

The reduced OEP model is executed using the critical time periods identified by CPS method and the candidate nodes from CBS method to determine the optimal BESS sizing and placement. The resulting BESS solution is then used to verify the voltage violation issue across the entire year. If validation fails, the planning window is expanded through the CPS method and the BESS planning problem is solved again. If the verification is successful, the results are finalized.

## IV. Generate Stochastic EV Charging Scenarios

The annual EV charging profiles used in this study are generated using our previously developed data-driven charging profile estimation framework [30]. The main steps required to construct the planning scenarios are summarized below. Based on whether EV charger is registered, the utility Center Point Energy serving the Greater Houston region, defines two residential user types: (i) non-EV users without any EV charger, and (ii) EV user with at least one EV charger. This work is based on realistic electricity consumption data at the user level; however, there are no dedicated EV charging data available. To address this challenge, we develop the following strategy to estimate EV charging profiles.

### *A. Extract EV charging curve*

The EV charging load is extracted from the total home-wide consumption data using the proposed multi-stage processing methodology. A reference non-EV user baseline signal is first established through averaging and normalization. The composite EV user signal is then scaled to match the amplitude of this reference. For accurate subtraction, the two signals are phase-aligned by synchronizing their 95th percentile troughs. The charging load signal is then isolated by subtracting the reference from the scaled composite signal. Finally, an inverse transformation restores the original amplitude of the extracted EV charging load. They are illustrated in Fig. 3.

From a dataset of 562,206 data points from different EV users, we identified 1,616 distinct charging events. A charging event is defined as a period where average power consumption continuously exceeds 4 kW for at least two hours or 7.2 kW for at least one hour, where average power is calculated based on the charging energy and duration of each charging event. The statistical characteristics of these events, such as charging duration, amount, start times, and end times, are compiled into histograms, and kernel density estimation is employed to create accurate, data-driven probabilistic models for analysis.

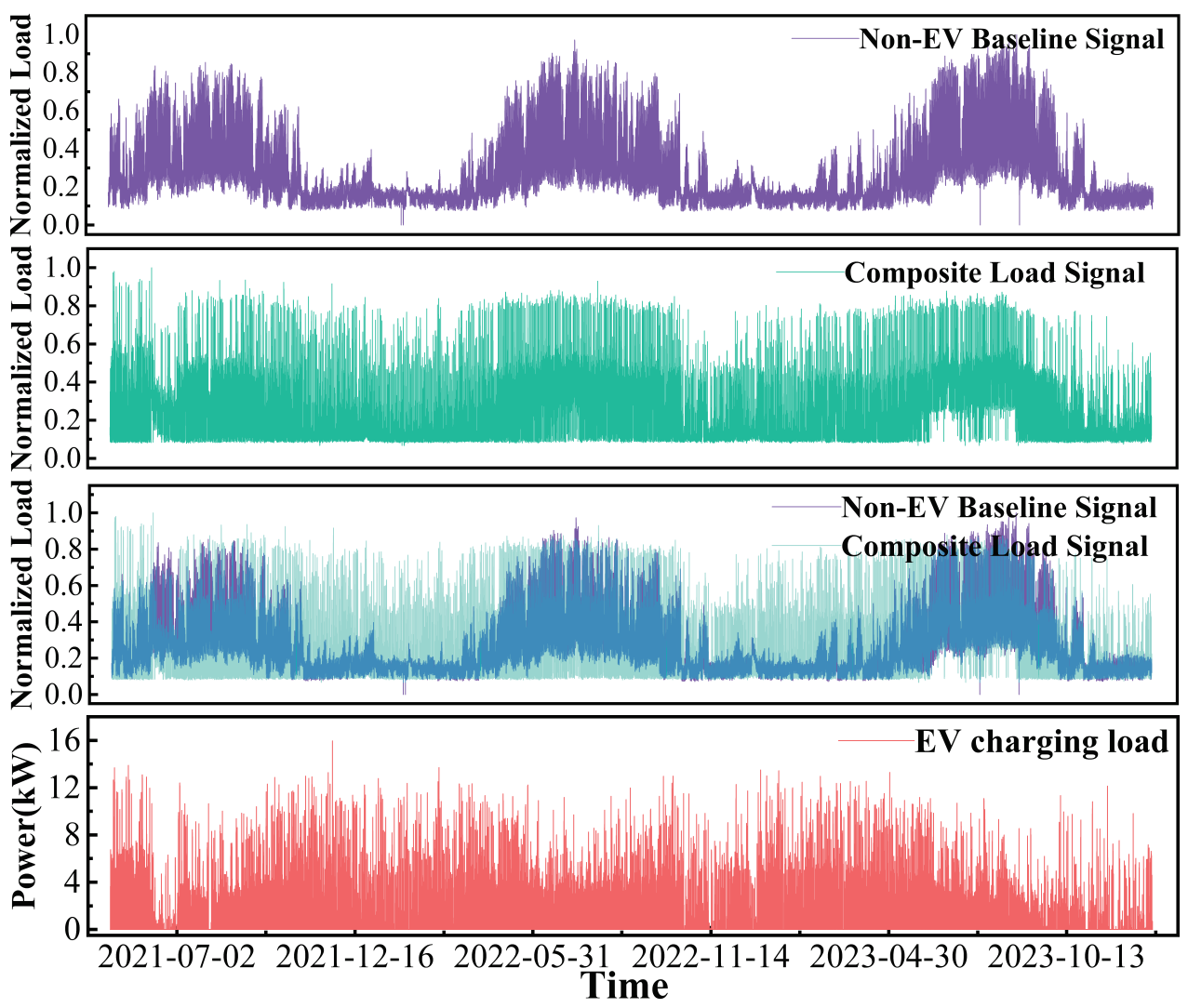

Fig. 3. (a) The graph of non-EV baseline signal, (b) graph of the composite load signal, (c) synchronization of non-EV baseline signal and composite load signal, (d) diagram of EV charging load curve.

### *B. EV charging curve probability distribution function*

Violin plots are used to visualize the probability distributions for charging capacity, as well as for the start and end times of charging events. As shown in Fig. 4, most charging events involve less than 30 kWh, peaking at approximately 10 kWh. Charging start times are mainly concentrated from 19:00 to 23:00, with end times mainly occurring between 21:00 and 2:00.

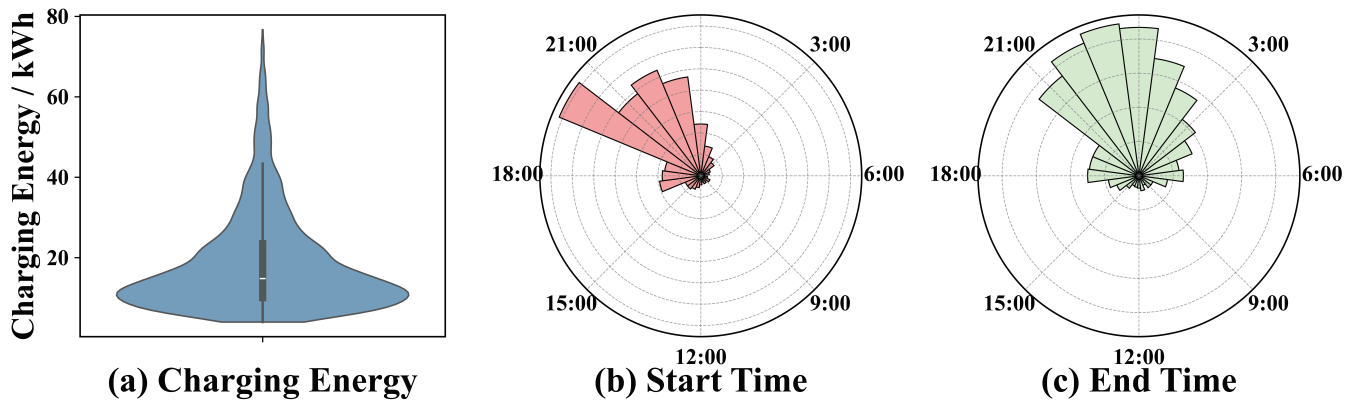

Fig. 4. (a) The diagram of charging energy distribution, (b) graph of the start time distribution, (c) graph of the end time distribution.

### *C. Random scenarios based on Monte Carlo method*

To avoid unrealistic combinations of charging duration and energy, such as excessively high or low charging power, a joint probabilistic model of charging duration and energy is employed. Based on the sampled values, the average charging power $P_{avg}$ is subsequently calculated. Events with the $P_{avg}$ below 4 kW are classified as low-power charging events, typically corresponding to EVs with a high SoC. Events with the $P_{avg}$ above 7.2 kW are categorized as the high-power charging events. All remaining events are treated as normal charging events. Based on the charging events setting, a Monte Carlo simulation is conducted to generate 1,200 annual charging scenarios, representing Level-2 chargers used by three types of EVs. A daily charging probability of 90% is assumed. The resulting charging energy profiles serve as the input for subsequent analysis and evaluation.

## V. Case Studies

The 240-bus test system [31] is derived from a real-world distribution network and includes a whole year of the corresponding electricity consumption data, detailing usage from both residential and industrial customers. Projected feeder demand is constructed by combining the baseline load trajectories with increasing levels of residential EV charging. The resulting impacts on the load curves in the distribution network are summarized in Table 2.

Table 2. Projected Growth of Electricity Consumption and EV Charging Load.

| *Time (year)* | *Hourly Load (kw)* | *Growth Factor* | *Charging Load (kw)* | *EV Penetration* |
|---|---|---|---|---|
| *0* | 6.68 | 1.0 | 0.23 | 10% |
| *5* | 7.35 | 1.1 | 0.68 | 30% |
| *10* | 8.02 | 1.2 | 1.36 | 60% |
| *15* | 8.69 | 1.3 | 2.26 | 100% |

To create robust test cases, load profiles incorporating EV charging load are stochastically sampled from a 100% EV penetration scenario. These profiles are then allocated to the standard IEEE 33-bus and 69-bus test systems to analyze voltage regulation challenges under high EV penetration. The optimization was conducted using Gurobi Optimizer v12.0.0, with a MIPGap of 0.001. The computational environment utilized a 12th Gen Intel® Core™ i7-12700 CPU. The model was implemented in Python 3.7.

## *A. Results of the IEEE 33-bus system*

By conducting the VVA analysis over the entire year, the 33-bus system experienced 6,975 instances of voltage violations across 17 distinct nodes, with the minimum nodal voltage dropping to 0.87p.u. at bus 18. The critical time period selected by the CPS is from June 9 to June 15.

The CPS method substantially reduces the temporal problem size and enables the formulation to be solved within the tested computational setting. To validate these two characteristics of the PSR strategy, we compared the optimization results of four distinct OEP models. These included (i) the full OEP model, which employed a full-year planning horizon and considered all violation nodes as candidates; (ii) the OEP model with CPS, which restricted the planning horizon to the critical period while retaining all violation nodes as candidates; (iii) the OEP model with CBS, which maintained a full-year planning horizon but used a reduced set of candidate nodes; and (iv) the OEP model with CPS and CBS, which combined both approaches by focusing on the same critical time period and using the reduced set of candidates. The specific planning and validation results for these four models are detailed in Table 3.

Table 3. Performance Comparison of Four OEP Models.

| Model | Time (s) | Best objective | Gap | Validation |
|---|---|---|---|---|
| *Full OEP* | 82233.34 | No solution | / | N/A |
| *OEP with CPS* | 11.16 | $0.684million | 0.0380% | Pass |
| *OEP with CBS* | 18071.62 | $0.685million | 0.0687% | Pass |
| *Reduced OEP* | 7.34 | $0.684million | 0.0394% | Pass |

Note: 'No solution' denotes the simulation is out of memory.

Based on the Table 3, the analysis clearly demonstrates the critical role of the PSR strategy in OEP model. The full OEP formulation did not produce a feasible incumbent before the computation terminated because of insufficient memory. In contrast, the CPS-based formulation was solved in 11.16 s and passed full-year validation, showing that temporal reduction enabled the problem to be solved within the tested computational setting. In contrast, all PSR-enabled models passed validation and delivered comparable best objective values about $0.684 million. The primary benefit of the PSR strategy lies in acceleration: the OEP model with CPS and CBS was the fastest, solving the problem in just 7.34 seconds, thereby validating both the CPS and CBS methods for accelerating the solution process.

To evaluate the effectiveness of the proposed strategy for selecting candidate nodes, we compared its performance against two established approaches under the same critical time period: (i) an Exhaustive method, serving as a comprehensive baseline by considering all nodes with recorded violations; and (ii) a Rank-based method that selects top-ranked candidate nodes based on their violation frequency.

Table 4. Candidate Nodes Selection and Optimized BESS Installation Capacities in the IEEE 33-bus System.

| *Method* | *Candidate Nodes* | *Optimized BESS Buses and Capacities (kWh)* |
|---|---|---|
| *Exhaustive* | 8, 9, 10, 11, 12, 13, 14, 15, 16, 17, 18, 28, 29, 30, 31, 32, 33. | 13: 143.22, 15: 843.85, 16: 324.94, 17: 244.38, 18: 495.16, 33: 229.92. |
| *Rank-based* | 10, 11, 12, 13, 14, 15, 16, 17, 18, 31, 32, 33. | 13: 143.18, 15: 843.70, 16: 402.06, 18: 662.70, 33: 229.90. |
| *PSR* | 9, 12, 13, 15, 16, 17, 18, 30, 31, 33. | 15: 986.94, 16: 401.96, 18: 662.82, 33: 229.87. |

Table 4 delineates the candidate nodes identified through three distinct methodologies and presents the corresponding BESS optimization results, where the maximum BESS capacity was limited to 1 MWh or 1,000 kWh.

The total installed capacity of the BESS optimized across all three candidate nodes selection methodologies was found to be almost identical. As demonstrated in Fig. 5, each method effectively mitigates the voltage violation problem, resulting in a similar voltage distribution.

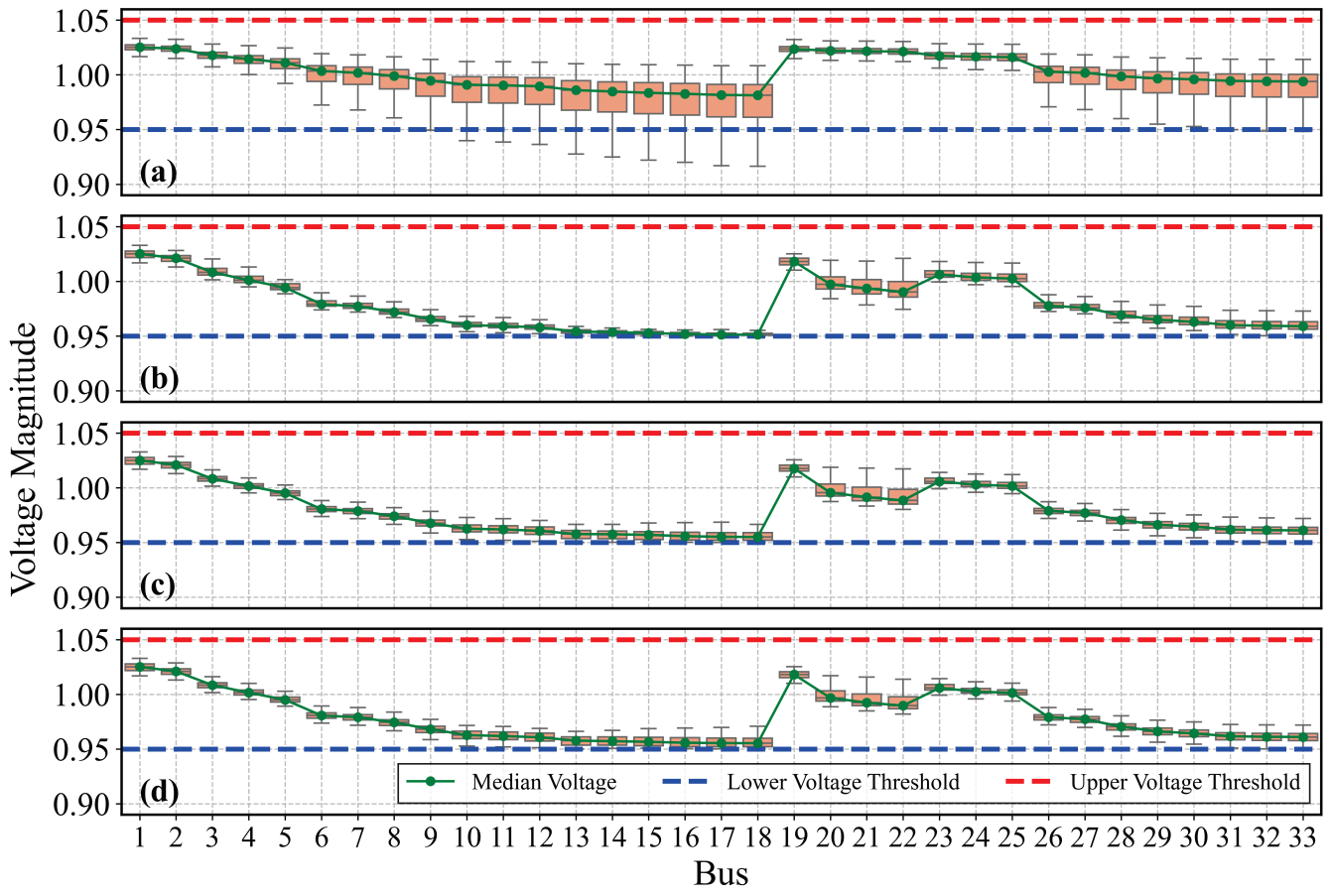

Fig. 5. Box plots of voltage distribution in IEEE 33-bus system: (a) voltage distribution without installing BESS, (b) voltage distribution with BESS based on *Exhaustive* method, (c) voltage distribution with BESS based on *Rank-based* method, (d) voltage distribution with BESS based on *PSR* strategy.

Table 5. Optimization Performance of Candidate Nodes Selection Methods in Planning and Validation Phases of the IEEE 33-bus system.

| Method | Time (s) | Best objective | Validation |
|---|---|---|---|
| *Exhaustive* | 11.16 | $0.684million | Pass |
| *Rank-based* | 8.58 | $0.684million | Pass |
| *PSR* | 7.34 | $0.684million | Pass |

As shown in Table 5, the proposed PSR strategy yielded an objective value of $0.684 million, which was identical to the optimal solution obtained by the Exhaustive method. The PSR solution matches the exhaustive benchmark in this case within the reported solver tolerance. All methods delivered high-quality results, with optimality gaps staying below 0.1% across all test systems. For the 33-bus system, the gap for the Exhaustive method was 0.0380%, the gap for the Rank-based

method was 0.0344%, and the gap for the PSR method was 0.0394%. This performance also extended to the larger 69-bus and 240-bus systems, confirming that all reported solutions remain close to the corresponding exhaustive benchmarks.

### *B. Results of the IEEE 69-bus system*

For the 69-bus system, the VVA results revealed 19,984 voltage violation infringements. The minimum nodal voltage observed was 0.88p.u. at bus 27, with violations distributed across 26 distinct buses. The critical time period was identified as July 15 to July 21. Table 6 shows the candidate nodes chosen by three different methods, including the optimized installation locations and capacities.

Table 6. Candidate Nodes Selection and Optimized BESS Installation Locations and Capacities in the IEEE 69-bus System.

| *Method* | *Candidate Nodes* | *Optimized BESS Locations and Capacities (kWh)* |
|---|---|---|
| *Exhaustive* | 12, 13, 14, 15, 16, 17, 18, 19, 20, 21, 22, 23, 24, 25, 26, 27, 58, 59, 60, 61, 62, 63, 64, 65, 68, 69 | 21: 24.490, 22: 1000.0, 23: 1000.0, 24: 1000.0, 25: 1000.0, 26: 1000.0, 27: 687.30, 64: 64.450, 65: 390.88. |
| *Rank-based* | 15, 16, 17, 18, 19, 20, 21, 22, 23, 24, 25, 26, 27 | 21: 1000.0, 22: 1000.0, 23: 1000.0, 24: 1000.0, 25: 1000.0, 26: 1000.0, 27: 955.62. |
| *PSR* | 16, 18, 19, 20, 21, 23, 24, 25, 26, 27, 65 | 21: 1000.0, 23: 1000.0, 24: 1000.0, 25: 1000.0, 26: 1000.0, 27: 712.71, 65: 455.35. |

Fig. 6 shows that the four sets of voltage profiles, without BESS and with BESS solutions. It indicates the three candidate nodes selection methods have very similar performance in terms of voltage regulation and violation mitigations; the proposed PSR strategy is the fastest as shown in Table 7, while it can achieve the optimal solution.

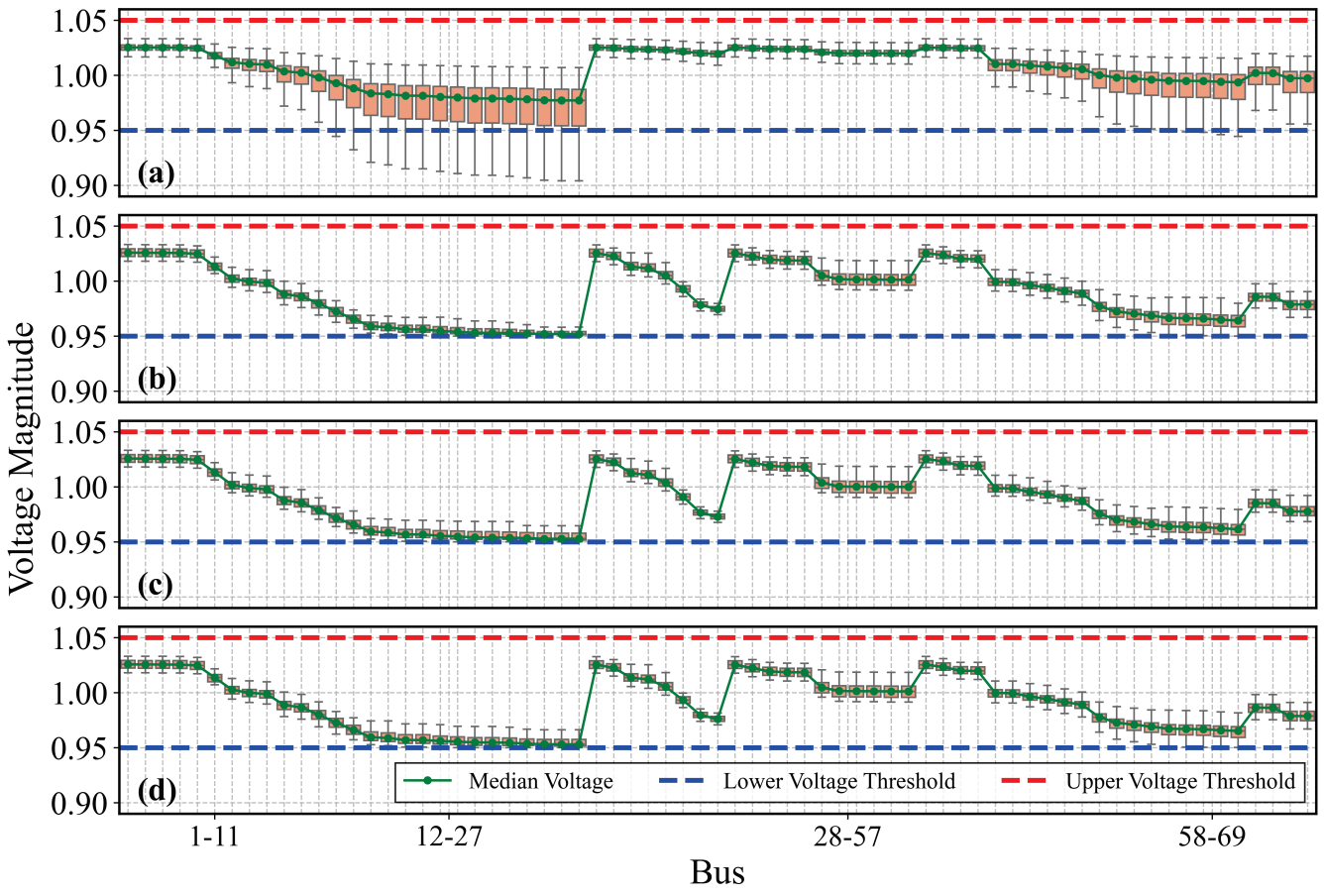

Fig. 6. Box plots of voltage distribution in IEEE 69-bus System: (a) voltage distribution without installing BESS, (b) voltage distribution with BESS based on *Exhaustive* method, (c) voltage distribution with BESS based on *Rank-based* method, (d) voltage distribution with BESS based on *PSR* strategy.

Table 7. Optimization Performance of Candidate Nodes Selection Methods in Planning and Validation Phases of the IEEE 69-bus system.

| Method | Time (s) | Best objective | Validation |
|---|---|---|---|
| *Exhaustive* | 1705.44 | $1.850million | Pass |
| *Rank-based* | 82.63 | $2.087million | Pass |
| *PSR* | 12.93 | $1.850million | Pass |

Table 7 shows the proposed strategy provided an objective value of $1.850 million, which again matched the optimal solution from the Exhaustive method. The PSR solution matches the exhaustive benchmark for this case while reducing the solution time by 99.2%. In contrast, the Rank-based method, while faster than the exhaustive search, yielded a suboptimal objective value for this case, highlighting a key advantage of the PSR strategy.

### *C. Results of the 240-bus system*

The 240-bus system is a fully monitored, radial distribution network modeled after a real-world Midwestern US utility, which uses a 69 kV substation to step down voltage and supply customers through three feeders. According to data of an entire year of 2017 at one-hour resolution, the load data covers 831 residential users across three feeders [31]. To focus on the balanced steady-state planning problem, the 240-bus system is transformed into a balanced network through replacing single-phase line configurations with equivalent three-phase counterparts. The base apparent power is set to 10 MVA, and the base voltage is 13.8 kV. Line current constraints are defined based on conductor configuration data.

Table 8. Candidate Nodes Selection and Optimized BESS Installation Capacities in the 240-bus System.

| *Method* | *Candidate Nodes* | *BESS Locations and Capacities (kWh)* |
|---|---|---|
| *Exhaustive* | 3031~3034, 3037~3039, 3051~3162. | 3112: 450.01, 3122: 466.08, 3127: 315.79, 3128: 1000.0, 3129: 1000.0, 3130: 978.13, 3131: 298.54, 3134: 810.90, 3135: 1000.0, 3136: 1000.0, 3137: 297.12, 3138: 737.95, 3139: 11.080, 3146: 88.090, 3147: 1000.0, 3153: 291.47, 3154: 189.64, 3155: 604.20, 3159: 351.36, 3160: 164.24, 3161: 592.42, 3162: 493.97. |
| *Rank-based* | 3131, 3130, 3129, 3128, 3127, 3126, 3136, 3125, 3135, 3122, 3138, 3139, 3134, 3121, 3124, 3137, 3120, 3123, 3133, 3132, 3119, 3118, 3162, 3161, 3160 | 3131: 325.32, 3130: 999.95, 3129: 1000.0, 3128: 1000.0, 3127: 1000.0, 3136: 1000.0, 3135: 1000.0, 3122: 1000.0, 3139: 1000.0, 3134: 1000.0, 3162: 1000.0, 3161: 1000.0, 3160: 1000.0. |
| *PSR* | 3112, 3115, 3122, 3124, 3126, 3128, 3131, 3134, 3136, 3137, 3139, 3143, 3147, 3155, 3162 | 3112: 134.46, 3122: 678.09, 3124: 1000.0, 3126: 1000.0, 3128: 1000.0, 3131: 1000.0, 3134: 999.97, 3136: 999.99, 3137: 684.50, 3139: 709.52, 3143: 1000.0, 3147: 1000.0, 3155: 1000.0, 3162: 1000.0. |

As for the VVA result, 2,093,640 nodal voltage values and corresponding line flow data are obtained; 118 nodes in feeder C experienced a total of 35,177 voltage dip in 620 different time periods. The lowest recorded nodal voltage was 0.91p.u., occurring at bus 3131. The critical time period was determined from July 3 to July 9. Table 8 provides the candidate nodes chosen by three methods, including the optimized installation capacities. As presented in Fig. 7, three methods effectively resolve the voltage violation issues. During the solution implementation period, the BESS operates between 10% and 90% SoC range, with a starting and ending point at 50%.

Table 9. Optimization Performance of Candidate Nodes Selection Methods in Planning and Validation Phases of the 240-bus system.

| Method | Time (s) | Best objective | Validation |
|---|---|---|---|
| *Exhaustive* | 2083.40 | $3.642million | Pass |
| *Rank-based* | 586.09 | $3.698million | Pass |
| *PSR* | 40.95 | $3.662million | Pass |

PSR reduces the solution time from 2,083.40 s to 40.95 s, while the investment objective differs from the exhaustive benchmark by 0.55%. The objective value obtained by the strategy was $3.662 million, deviating from the benchmark value ($3.642 million) by 0.55%. The result demonstrates that PSR can substantially reduce computational effort when full candidate searches become expensive.

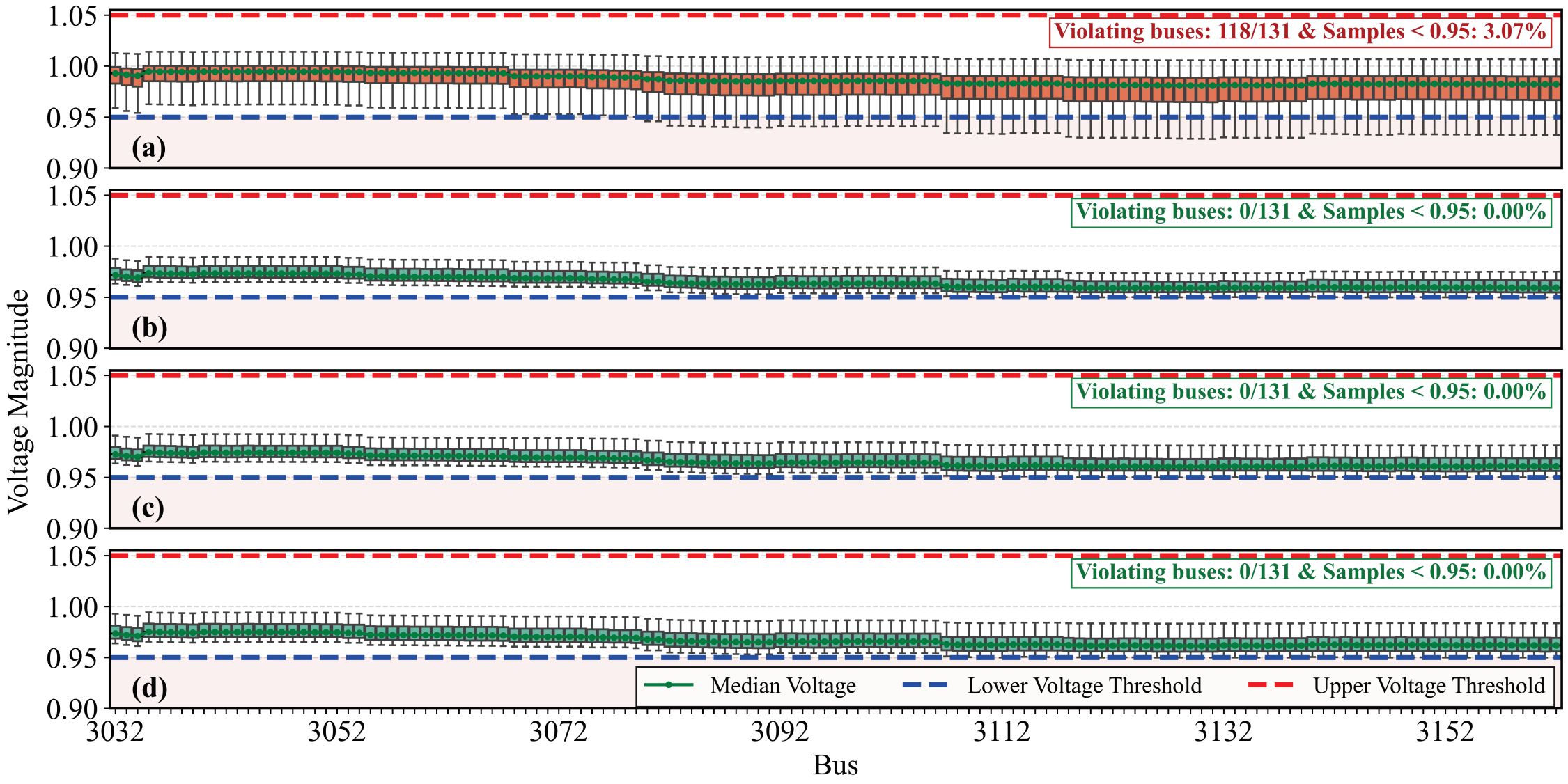

Fig. 7. Box plots of voltage distribution in the 240-bus system: (a) voltage distribution without installing BESS, (b) voltage distribution with BESS based on the *Exhaustive* method, (c) voltage distribution with BESS based on the *Rank-based* method, (d) voltage distribution with BESS based on the *PSR* strategy.

## VI. Economic and Operational Benefits of BESS Integration

This section evaluates the economic feasibility and operational benefits of BESS integration over a 15-year planning horizon. This study utilizes real-world annual electricity price data to assess its impact on system operation considering the time-of-use pricing. A comparative analysis of electricity costs is performed for two scenarios: the system without installing BESS and the system with BESS optimally placed at candidate nodes identified by the proposed PSR strategy in 33-bus and 69-bus and 240-bus systems.

Furthermore, this analysis aims to determine whether BESS deployment can effectively balance load and electricity price fluctuations to achieve overall lower electricity costs, in addition to its primary function of resolving voltage violation issues. Meanwhile, the objective of the model is updated as:

$$objective = min\left(\sum_{s\in\mathcal{G},\forall t\in\mathcal{T}} (c_t^{operation} \cdot P_s^t)\right) \quad (40)$$

The results in Table 10 and Table 11 demonstrate that the integration of the BESS yields substantial and scalable economic benefits and energy reductions. The deployment of BESS provides consistent and significant reductions in annual operational costs. The magnitude of this cost-saving scales proportionally with the size of the system. Specifically, the average annual savings are approximately $0.14 million for the 33-bus system, $0.33 million for the 69-bus system, and $0.76 million for the 240-bus system.

The BESS integration provides effective mitigation of annual energy losses. The absolute energy savings increase substantially with both load growth over time and system scale. For instance, in the 33-bus system, the annual loss reduction grew from 18.36 MWh to 92.02 MWh. This trend is more pronounced in larger systems: the 69-bus system shown an increase in reductions from 30.49 MWh to 174.15 MWh, while the 240-bus system's loss reduction expanded from 104.98 MWh to 438.2 MWh. These results indicate additional economic value from the planned BESS installations beyond voltage mitigation.

Table 10. Annual Cost and Economic Benefits with BESS Integration.

| **33-bus system** | | | | |
|---|---|---|---|---|
| *EV Penetration* | *w.o. BESS* | *w. BESS* | *Savings* | |
| 10% | $0.97 M | $0.83 M | $0.14 M | 14.43% |
| 30% | $1.17 M | $1.03 M | $0.14 M | 11.97% |
| 60% | $1.44 M | $1.29 M | $0.15 M | 10.42% |
| 100% | $1.76 M | $1.61 M | $0.15 M | 8.52% |
| **69-bus system** | | | | |
| *EV penetration* | *w.o. BESS* | *w. BESS* | *Savings* | |
| 10% | $2.08M | $1.76 M | $0.32 M | 15.38% |
| 30% | $2.52 M | $2.20 M | $0.32 M | 12.70% |
| 60% | $3.10 M | $2.76 M | $0.34 M | 10.97% |
| 100% | $3.81 M | $3.46 M | $0.35 M | 9.19% |
| **240-bus system** | | | | |
| *EV penetration* | *w.o. BESS* | *w. BESS* | *Savings* | |
| 10% | $6.15 M | $5.40 M | $0.75 M | 12.20% |
| 30% | $7.16 M | $6.41 M | $0.75 M | 10.47% |
| 60% | $8.39 M | $7.63 M | $0.76 M | 9.06% |
| 100% | $9.85 M | $9.07 M | $0.78 M | 7.92% |

Table 11. Annual Energy Loss and Reductions with BESS Integration.

| **33-bus system** | | | | |
|---|---|---|---|---|
| *EV penetration* | *w.o. BESS* | *w. BESS* | *Reductions* | |
| 10% | 150.41MWh | 132.05MWh | 18.36MWh | 12.21% |
| 30% | 227.51MWh | 198.96MWh | 28.55MWh | 12.55% |
| 60% | 360.46MWh | 310.31MWh | 50.15MWh | 13.91% |
| 100% | 582.88MWh | 490.86MWh | 92.02MWh | 15.79% |
| **69-bus system** | | | | |
| *EV penetration* | *w.o. BESS* | *w. BESS* | *Reductions* | |
| 10% | 259.06MWh | 228.57MWh | 30.49 MWh | 11.77% |
| 30% | 397.49MWh | 347.99MWh | 49.50 MWh | 12.45% |
| 60% | 639.90MWh | 550.49MWh | 89.41 MWh | 13.97% |
| 100% | 1050.7MWh | 876.55MWh | 174.15MWh | 16.57% |
| **240-bus system** | | | | |
| *EV penetration* | *w.o. BESS* | *w. BESS* | *Reductions* | |
| 10% | 1010.9MWh | 905.92MWh | 104.98MWh | 10.38% |
| 30% | 1427.7MWh | 1276.6MWh | 151.10MWh | 10.58% |
| 60% | 2084.6MWh | 1834.8MWh | 249.80MWh | 11.98% |
| 100% | 3107.5MWh | 2669.3MWh | 438.20MWh | 14.10% |

Note: 'w.o.' denotes 'without', and 'w.' denotes 'with'.

## VII. Conclusions

The increasing penetration of EV charging can create persistent voltage violations in distribution networks, requiring additional grid support as charging demand grows. This paper developed a proactive voltage management (PVM) framework that integrates stochastic EV charging scenario generation, annual voltage violation assessment, and BESS planning. The EV charging profiles were generated through Monte Carlo simulation using probability distributions fitted to real charging events, providing realistic annual charging conditions for the planning study.

To improve the computational tractability of BESS planning, a PSR strategy was developed to reduce both the temporal horizon and candidate siting space. The CPS method identifies high stress periods from annual voltage violation records, while the CBS method reduces the BESS candidate set using voltage sensitivity, violation characteristics, clustering, and network information. The CPS and CBS methods reduce the chronological horizon and BESS candidate set before investment optimization. The resulting installations are then checked against the complete annual operating trajectory.

Case studies on the IEEE 33-bus, IEEE 69-bus, and 240-bus systems show that the proposed approach substantially reduces the computational burden while retaining planning performance comparable to the benchmark methods. The optimized BESS deployments eliminate the identified voltage violations in the tested cases and also reduce electricity purchase costs under time-of-use tariffs. These results demonstrate the effectiveness of combining annual EV impact assessment, planning space reduction, and BESS investment optimization for distribution networks with increasing EV charging demand.